\documentstyle[11pt,aasms4]{article}
\begin{document}

\title{Long-term Variability Properties and 
Periodicity Analysis for Blazars}

\author{J.H. Fan}

\affil{
1. Center for Astrophysics, Guangzhou Normal University, Guangzhou 510400,
 China, e-mail: jhfan@guangztc.edu.cn\\
2. Chinese Academy of Sciences-Peking
University Joint Beijing
Astrophysical Center (CAS-PKU.BAC), Beijing, China}

\date{Received <date>;accepted<date>}

\begin{abstract}

In this paper, the compiled long-term optical and infrared measurements
of some blazars are used to analysis the variation properties and the
optical data are used to search for 
periodicity evidence in the lightcurve by means of the Jurkevich
technique and the discrete correlation function (DCF) method.
Some explanations have been discussed.


\end{abstract}

\section{Introduction}

The nature of the central regions of quasars and other Active
Galactic Nuclei (AGNs) is still an open problem. The study of AGN
optical variability can yield valuable information about the
mechanisms operating in these sources, with important implications
for quasar modeling (see, for instance, Fan et al. 1998). Some
particular objects have been claimed to display periodicity in
their lightcurves over a variety of timescales (e.g. Jurkevich
1971;  Hargen-Thorn et al. 1987; Sillanpaa et al. 1988; 
Webb et al. 1988;
Babadzhanyants \& Belokon 1991; 
Kidger et al. 1992; 
Liu et al. 1995; 
Marchenko et al. 1996;
Fan et al. 1997, 1998, 
Fan \& Su 1998; 
Fan 1999a; 
Zhang et al. 1998;
Fan \& Lin 1999a; 
Fan et al. 1999a; 
Lainela et al. 1999;
Lin 1999; 
Su 2000 and references therein), but, in general, the clear
identification of periodic
behaviour has been very elusive due to the complexity of the
optical lightcurves and the lack of databases large enough as to
provide an adequate sampling over large periods.

Blazars are an extremely subclass of AGNs, characterized as rapid
and high amplitude variations, high and variable polarization,
superluminal radio components, nonthermal continuum.  Blazars have
been monitored for a long time, compilations are available for
optical (Fan \& Lin 1999b,c) and infrared (Fan \& Lin 1999d; Fan 1999b)
bands. The infared data are available in
http://xxx.lanl.gov/astro-ph/9908104 and 
http://xxx.lanl.gov/astro-ph/9910269. The
long-term measurements make it possible for one to search
for periodicities from the light curves and to discuss the long-term
variability properties.
Because the characteristic of the
astronomical measurements, we adopted the Jurkevich (Jerkevich 1971) and
DCF (Discrete Correlation Function) (see Fan \& Lin 1999a) methods
to the available data in searching for the periods. 
In the second 
section we will give the variation properties, in the third
section, we present
the results and finally in section 4, we give some discussions.

\section{Variation Property}

From the complication (Fan et al. 1999b,c,d; Fan 1999b), we found that the
largest variations are comparable in both the optical and infrared bands.
Largest variations at different wavelengths increase with decreasing
wavelength. The variations are also correlated with the highest 
observed optical polarization, which can be explained using the
beaming model. Some
objects show that the spectra flatten when the sources
brighten while the opposite cases are for some others. For the largest
infrared and optical variations, we listed them in Table 1.

\section{ Periodicity Analysis Methods}

\subsection{Jurkevich Method}

 The Jurkevich method (
 Jurkevich 1971, a
 lso see Kidger et al. 1992; 
 Fan et al. 1998a; 
 Fan 1999a)
 is based on
 the expected mean square deviation and it is less inclined to generate
 spurious periodicity than the Fourier analysis used by other authors (e.g.
 Fan et al. 1997). It tests a run of trial
 periods around which the data are folded. All data are assigned to $m$
 groups according to their phases around each trial period. The variance
 $V_i^2$ for each group and the sum $V_m^2$ of all groups are then computed.
 If a trial period equals the true one, then $V_m^2$ reaches its minimum.
 So, a ``good'' period will give a much reduced variance relative to those
 given by other false trial periods and with almost constant values.
 A further test is the relationship
 between the depth of the minimum and the noise in the ``flat'' section of
 the $V_m^2$ curve close to
 the adopted period. If the absolute value of the relative change of the
 minimum to the ``flat'' section is large enough as compared with the
 standard error of this ``flat'' section (say, five times), the
 periodicity in the data can be considered as significant and the minimum
 as highly reliable (Fan et al. 1998a).

\subsection{DCF Method}

 The DCF (Discrete Correlation Function) method is intended for analyses
 of the correlation of two data set. It is described in detail by
 Edelson \& Krolik (1988) (also see Fan et al. 1998b). This method can
 indicate the
 correlation of two variable temporal series with a time lag, and can
 be applied to the periodicity analysis of a unique temporal data set. If
 there is
 a period, $P$, in the lightcurve, then the DCF should show
 clearly whether the data set is correlated with itself with time lags of
 $\tau$ = 0 and $\tau$ = $P$ (see Fan \& Lin 1999a). 
 We have implemented the method as  follows.

 Firstly, we have calculated the set of
 unbinned correlation (UDCF) between data points in the two data streams
 $a$ and $b$, i.e.

\begin{equation}
 {UDCF_{ij}}={\frac{ (a_{i}- \bar{a}) \times (b_{j}- \bar{b})}{\sqrt{\sigma_{a}^2 \times
 \sigma_{b}^2}}},
\end{equation}
 where $a_{i}$ and $ b_{j}$ are points in the data sets, $\bar{a}$ and $\bar{b}$
 are
 the average values of the data sets, and $\sigma_{a}$ and $\sigma_{b}$ are
 the corresponding standard deviations. Secondly, we have averaged the
 points sharing the same time lag by
 binning the $UDCF_{ij}$ in suitably sized time-bins in order to get the
 $DCF$ for each time lag $\tau$:

\begin{equation}
 {DCF(\tau)}=\frac{1}{M}\Sigma \;UDCF_{ij}(\tau),
\end{equation}
 where $M$ is the total number of pairs. The standard error for each bin is

\begin{equation}
\sigma (\tau) =\frac{1}{M-1} \{ \Sigma\; [ UDCF_{ij}-DCF(\tau)
]^{2} \}^{0.5}.
\end{equation}

\section{Results and Discussion}

 Very recently, we compiled four  data bases for BL Lac objects 
 and OVV/HPQs in the optical and infrared bands
 (Fan \& Lin 1999b,c,d; Fan 1999b). The data bases can be used to 
 discuss the long-term variability properties, the correlated variations
 and to search for periodicity signatures in the light curves.
 When the both methods are adopted to those optical
 data, the both methods give consistent results.  The results are presented
 in Table 2.  It is interesting to notice that there is a common period 
 of $\sim$ 1.0 year, which is likely from the effect of the Sun on the
 measurements. The detail analysis is presented in the paper by Fan et al.
 (1999c) 

 For the long-term periodicity of variations, there are several explanations:
 the double black hole model, the hectic jet model, the slim disk model,
 and the effect of external perturbations to the accretion disk. (e.g.
 Sillanpaa et al. 1988; Meyer \& Meyer-Hofmeister 1984; Horiuchi \& Kato 1990;
 Abraham \& Romero 1999; Fan et al. 1997, 1998a, 1999; Villata \&
 Raiteri 1999; Romero et al. 2000).
 
\begin{acknowledgments}
{This work has been supported by the National PanDeng Project of
China and the National Natural Scientific Foundation of China.
I thank Prof. K.S. Cheng to provide me with the opportunity
to present this work and Prof. R.G. Lin and C.Y. Su for their help.}
\end{acknowledgments}

\newpage
\begin{small}
\begin{table*}
\caption[]{The largest variations of  Blazars}
\begin{tabular}{lccc||lccc}
\noalign{\smallskip}
\hline
\noalign{\smallskip}
 Name     & $\Delta m_{opt}$ &$\Delta m_{IR}$ & $P_{opt} (\%)$ &
 Name     & $\Delta m_{opt}$ &$\Delta m_{IR}$ & $P_{opt} (\%)$ \\
 (1)      & (2)              & (3)            & (4)           &
 (1)      & (2)              & (3)            & (4)            \\ \hline 
0048-097 & 2.7 & 6.55 & 27.2  & 0109+244  & 3.07 & 1.58 & 17.3\\
0118-272  & 1.05 & 0.67  & 17.& 0138-097  & 1.52  & 1.69 & 29.3\\
0215+015  & 5.0 & 2.69  & 20. & 0219+428  & 2.0 & 1.61 & 18.0\\
0235+164  & 5.3 & 5.0  & 44.  & 0323+022  & 1.3 & 0.97 & 10.4 \\
0420-014  & 2.8 & 2.88 & 20.  & 0422+004  & 2.2 & 3.25 & 22.\\
0521-365  & 1.4 & 1.25 & 11.  & 0537-441  & 5.4 & 3.0  & 18.8\\
0716+714  & 5.0 &      & 29.0 & 0735+178  & 4.6 & 2.47 & 36.\\
0736+017  & 1.35 & 2.71 & 6.  & 0754+100  & 3.16 & 1.88 & 26.\\
0818-128  & 3.78 & 2.14 & 36. & 0823-223  & 1.41 &  2.32 & 11.\\
0829+046  & 3.58 & 2.15 & 12. & 0851+202 & 6.0 & 3.87 & 37.2\\
0912+297  & 2.25 & 2.27 & 13. & 1101+384 & 4.6 & 2.03 & 16.0\\
1144-379 & 1.92 & 3.46 & 8.5  & 1147+245 & 1.0 & 1.34 & 13.0\\
1156+295  & 5.0 & 4.47 & 28.  & 1215+303  & 3.3 & 1.05 & 14.\\
1219+285  & 3.13 & 2.46 & 10. & 1253-055  & 6.7 & 4.57 & 44.\\
1308+326  & 4.17 & 5.55 & 28. & 1418+546  & 4.8 & 1.59 & 24.\\
1510-089  & 5.4 & 1.25 & 14.  & 1514-241  & 3.0 & 2.64 & 8.0\\
1538+149  & 3.7 & 0.89 & 32.8 & 1641+395  & 3.0 & 3.16 & 35.\\
1652+398  & 1.30 & 2.37 & 7.0 & 1727+502  & 2.1  & 0.82 & 6.0\\
1749+096  & 2.7 & 2.21 & 32.  & 1807+698  & 2.0 & 1.02 & 12.\\
1921-293  & 2.64 & 3.06 & 17. & 2005-489  & 0.53 & 0.31 & 2.0 \\
2155-304  & 1.85 & 1.88 & 14.2& 2200+420  & 5.31 & 2.93 & 23.\\
2223-052  & 5.0 & 3.96 & 17.3 &2240-260  &   &  1.66 & 15.1\\
2251+158  & 2.5 & 1.57 & 19. &2254+074  & 3.27 & 1.36 & 21.\\
\hline
\end{tabular}
~~~~~~~~~~~~~~~~~~~~~~~~~~~~~~~~~~~~~~~~Notes to the table:\\
Col. 1 The name of the source; \\
Col. 2, The largest optical variation;\\
Col. 3 The largest infrared variation; \\
Col. 4, The largest optical polarization
\end{table*}
\end{small}

\begin{table*}
\caption[]{Investigation Results of Periodicity of 14 Blazars}
\begin{tabular}{lllll}
\noalign{\smallskip}
\hline
\noalign{\smallskip}
 Name       & 
$P_{1}~(f)$(year) & 
$P_{2}~(f)$(year) & 
$P_{3}~(f)$(year) & 
$P_{4}~~(f)$(year) \\
\hline 
0219+428  & 0.99    & 2.14            & 2.93 & 4.52  \\
0235+164  & 1.01    & 1.56            & 2.95$\pm$0.25 & 5.87$\pm$1.3 \\
0735+178  & 1.01    & 14.39$\pm$0.51  & 18.58$\pm$0.52 & 28.62$\pm$0.58\\
0754+100  & 1.00    & 17.85$\pm$1.3   & 24.7$\pm$0.7 &  \\
0851+202  & 1.01    & 5.53            &11.75$\pm$0.5  &  \\
1215+303  & 0.99    & 2.05$\pm$0.1    & 4.45$\pm$0.05 & 6.89$\pm$0.34\\
1219+285  & 0.99    & 1.97            & 9.0$\pm$0.1 & 14.84$\pm$1.55   \\
1226+023  & 0.98    & 2.00            & 13.5$\pm$0.2 & 22.5$\pm$.02  \\   
1253-055  & 1.2     & 7.1$\pm$0.4     &      & \\
1308+326  & 1.00    & 2.03            & 2.95 & 6.07$\pm$0.1  \\
1418+546  & 0.99    & 2.99            &           &           \\
1514-241  & 1.00    & 2.01            & 4.09 & 9.35$\pm$0.1 \\
1807+698  & 0.99    & 6.0             & 18.18 &                     \\
2155-304  &         & 4.16            & 7.0      & \\
2200+420  & 0.99    & 13.7$\pm$0.4    & 19.95$\pm$0.70  & 29.6$\pm$0.30 \\ 
\hline
\end{tabular}
~~~~~~~~~~~~~~~~~~~~~~~~~~~~~~~~~~~~~~~~~~Notes to the Table:\\
$P_{1}$, $P_{2}$, $P_{3}$, $P_{4}$ are the possible periods derived
from the light curves 
\end{table*}

\cite{}


\begin{thebibliography}{}
\bibitem[]{} Abraham, Z. \& Romero, G.E. 1999, A\&A, 344, 61, 1999
\bibitem[]{} Angione, R.J. et al., 1981, AJ, 86, 653
\bibitem[]{} Angione, R.J. \& Smith, H.J. 1985, AJ, 90, 2474
\bibitem[]{} Babadzhanyants, M.K. \& Belokon, E.T., 1991, in Variability
of Blazars
eds. E. Valtaoja \& M. Valtonen, Cambridge University Press, p384
\bibitem[]{} Fan J.H., Xie G.Z., Lin R.G. et al. 1997, A\&AS, 125, 525
\bibitem[]{} Fan J.H. \& Su C.Y. 1998, Ch.A\&A, 23, 22
\bibitem[]{} Fan J.H., Xie G.Z., Pecontal E., et al., 1998a, ApJ, 507,
173
\bibitem[]{} Fan J.H., Adam G. Xie G.Z. et al., 1998b, A\&AS, 136, 217
\bibitem[]{} Fan J.H., et al. 1999a, ASP Conf. Ser. Vol. 159, p99
\bibitem[]{} Fan J.H., Romero, G. \& Lin R.G.  1999b, submitted
\bibitem[]{} Fan J.H., et al. 1999c, A\&A, submitted
\bibitem[]{} Fan J.H., 1999a, MNRAS, 382, 1032
\bibitem[]{} Fan J.H., 1999b, astro-ph/9910269
\bibitem[]{} Fan J.H. \& Lin R.G. 1999a, A\&A, (accepted)
\bibitem[]{} Fan J.H. \& Lin R.G., 1999b, (preparation)
\bibitem[]{} Fan J.H. \& Lin R.G., 1999c, ApJ, (accepted)
\bibitem[]{} Fan J.H. \& Lin R.G., 1999d, ApJS, 121, 131, astro-ph/9908104
\bibitem[]{} Hagen-Thorn et al. 1987, Astron. Rep. 41, 1
\bibitem[]{} Honma, F. et al. 1991, PASJ, 43, 147
\bibitem[]{} Horiuchi, T. \& Kato, S. 1990, PASJ, 42, 661
\bibitem[]{} Hufnagel, B.R.  \& Bregman, J.N. 1992, ApJ, 386, 473
\bibitem[]{} Jurkevich, I. 1971, Ap\&SS, 13, 154
\bibitem[]{} Kidger, M.R., Takalo, L., Sillanpaa, A. 1992, A\&A, 264, 32
\bibitem[]{} Lainela M., et al. 1999, ApJ, 521, 561
\bibitem[]{} Lin R.G. 1999, Acta Astrophysica Sinica, submitted
\bibitem[]{} Liu F.K., Xie G.Z., Bai J.M. 1995, A\&A, 295, 1
\bibitem[]{} Marchenko S.G. et al. 1996, ASP Conf. Ser. 110, p105.
\bibitem[]{} Romero, G.E. et al. 2000, A\&A, submitted
\bibitem[]{} Sillanpaa, A., Mikkola, S., Valtaoja, L. 1991, A\&AS, 88,
225
\bibitem[]{} Sillanpaa, A., Haarala, S., Korhonen, T. 1988, A\&AS, 72,
347
\bibitem[]{} Su C.Y.  2000, Acta Astrophysica Sinica, (in press).
\bibitem[]{} Villata M., \& Raiteri C.M. 1999, A\&A, 347, 30
\bibitem[]{} Webb J.R. et al. 1988, AJ, 95, 374
\bibitem[]{} Zhang X., Xie G. Z., Bai J.M., 1998, A\&A, 330, 469
\end{thebibliography}
\end{document}